\documentstyle[psfig]{mnv2}
\newif\ifAMStwofonts
\AMStwofontstrue

\def\gsim{{\mathrel{\raise0.35ex\hbox{$\scriptstyle >$}\kern-0.6em 
\lower0.40ex\hbox{{$\scriptstyle \sim$}}}}}

\def\yr{\,{\rm yr}}
\def\kms{\mbox{km\,s$^{-1}$}}
\newcommand{\Mpc}{\, {\rm Mpc} }
\def\Msol{\mbox{M$_\odot$}}

\newcommand{\op}{Ly$\alpha$\ }

\title[High-redshift galaxies, their active  nuclei  and
central black holes]
{High-redshift galaxies, their active nuclei  and
central black holes}

\author[]
{Martin G. Haehnelt, Priyamvada Natarajan and  Martin J. Rees\\ 
Institute of Astronomy, Madingley Road, Cambridge CB3 0HA}

\pagerange{\pageref{firstpage}--\pageref{lastpage}}
\pubyear{1994}

\begin{document}

\maketitle

\label{firstpage}

\begin{abstract}
We demonstrate that the luminosity function of the recently 
detected population of actively star-forming galaxies
at redshift three and the  B-band QSO luminosity function at  
the same redshift can both be matched with the mass function of 
dark matter haloes predicted by standard variants of hierarchical 
cosmogonies for lifetimes of optically bright QSOs  anywhere in the range 
$10^{6}$ to $10^{8}\yr$.  There is a strong correlation between
the lifetime and the required degree of non-linearity 
in the relation between black hole and halo mass.  
We suggest that the mass of supermassive black 
holes may be limited by the back-reaction of the emitted energy 
on the accretion flow in a self-gravitating disc. This would imply     
a relation of black hole to halo mass of the form $M_{\rm bh} 
\propto v_{\rm halo}^5 \propto M_{\rm halo}^{5/3}$ 
and  a typical  duration of the optically bright 
QSO phase of  a few times $10^{7} \yr$. The  high integrated  
mass density of  black holes  inferred from  recent 
black hole mass estimates in nearby galaxies may indicate  
that the overall efficiency of supermassive black holes 
for producing blue light is smaller than previously assumed. 
We discuss three possible accretion modes with low optical 
emission efficiency:  
(i) accretion at far above the Eddington rate, 
(ii) accretion  obscured by dust,  and 
(iii) accretion below the critical rate leading to  an 
advection dominated accretion flow lasting for a Hubble time. 
We further  argue that accretion with low optical efficiency 
might be closely related to the  origin of the hard X-ray background 
and that the ionizing background might be progressively dominated by 
stars rather than QSOs at higher redshift.
\end{abstract}


\section[]{Introduction}
For three  decades optically bright QSOs were our only beacons 
in the high redshift universe. Soon after the discovery of QSOs it
became  clear that their comoving space density increases dramatically
up to redshift two to three. Larger systematic surveys confirmed early
suggestions that this evolution is reversed at even higher redshift and
established an epoch of optically bright QSO
activity with a peak around redshift $z$=2.5 (Schmidt, Schneider \&
Gunn 1994; Warren , Hewett \& Osmer 1994; Shaver et al.~1996; McMahon,
Irwin \& Hazard 1997). 
QSO activity is widely  believed to be the result of accretion 
onto super-massive black holes at the centre of galaxies 
(e.g. Rees 1984 and references therein) and a number of authors have
linked the change in QSO activity to a corresponding change in fuel
supply at the centre of the host galaxies (Cavaliere \& Szalay 1986;
Wandel 1991; Small \& Blandford 1992; Yi 1996).  
Haehnelt \& Rees (1993; HR93 henceforth; see also Efstathiou \& Rees 
1988)  recognized that the epoch of QSO activity coincides with 
the time when the first deep potential wells form in plausible variants of 
hierarchical cosmogonies.  

The past few years have seen dramatic improvements in detecting
ordinary galaxies and starbursts out to $z\ga$3, transforming 
our knowledge of galaxy and star formation in the high redshift
universe. There are also far more extensive data  on 
the demography of supermassive black holes in nearby galaxies, 
and on  low level activity of AGN both in the optical and the 
X-ray bands. We discuss  here some implications for  the formation and
evolution of active galactic nuclei, attempting to tie these lines of 
evidence together  in a consistent model. 

The paper is organized as follows. In section 2 we discuss the general 
framework used to relate the luminosity function of QSOs and
star-forming galaxies to the mass function of dark matter (DM) haloes. 
Section 3 focuses on constraints on the accretion history 
of supermassive black holes. Section 4 summarizes  observational implications
and Section 5 briefly discusses the respective role of stars and QSOs
for the re-ionization history of the inter-galactic medium. Section 6  
contains our conclusions.

\section[]{Star-formation and QSO activity at high redshift}

\subsection{The space density of dark matter haloes}

\begin{figure}
\centerline{
\hspace{0.0cm}\psfig{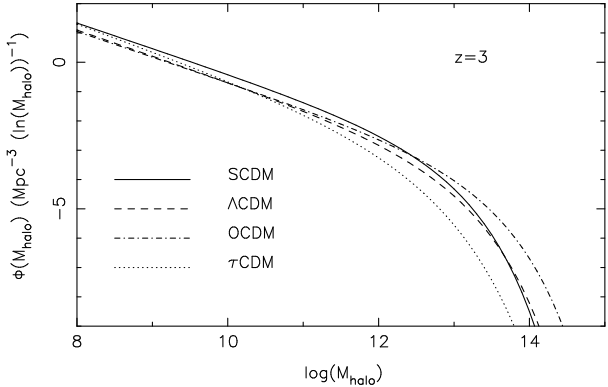}
}
\vspace{0.0cm}
\caption{Press-Schechter estimate of the space density of dark matter
haloes at $z$=3 for four different CDM variants chosen by Jenkins
et al.~(1997) to comply with observational constraints in the low
redshift universe on galaxy and cluster scales.
SCDM is a standard CDM model, $\Lambda$CDM and OCDM are 
open models with and without a cosmological constant and 
$\tau$CDM mimics a mixed dark matter model with a neutrino 
contribution to the matter content. 
Detailed parameters of the CDM variants are shown in Table 1. 
\label{fig_1}}
\end{figure}

It is widely  believed that the matter distribution in the universe 
is dominated by  dark matter and that the gravitational growth of 
density  fluctuations  created in the early universe is
responsible for the  formation of the observed structures.  
The strongest constraints on the fluctuation amplitude 
of the DM density field on small scales  comes from the observed 
space density of rich galaxy clusters which probe only slightly larger
scales than we are interested in here (White, Frenk \& Efstathiou 1993). 
We therefore take   the four variants of  cold dark matter (CDM)  
models chosen by Jenkins et al.~(1997) to comply with observational 
constraints in the low redshift universe on galaxy and cluster scales
as a representative sample of viable models (for parameters see Table 1).  

The Press-Schechter formalism 
gives simple but still  reasonably accurate estimates of the 
space density of DM haloes. In Fig. 1 we plot the mass function of DM
haloes at $z$=3 for the four models. As expected the mass function
is essentially a power-law at small masses and falls off exponentially
at the high mass end.  While at $z$=0, the mass functions for the 
different models are, by design, very similar, at $z$=3 the 
characteristic 
turnover mass in the mass function differs considerably between the
models.  In the next two sections we will explore 
the link between  star-forming galaxies and QSOs at high redshift, 
assuming that both populations trace the mass function of  
DM haloes. 

\begin{table}
\caption{The model parameter of the CDM variants explored:
$\sigma_8$ is the  {\it rms} linear overdensity in spheres of radius    
$8\,h^{-1}\,{\rm Mpc}$ and $\Gamma$ is the shape parameter 
as defined in the fitting  formula for CDM-like spectra by 
Efstathiou, Bond \& White (1992).}
\begin{tabular}{l|l|l|l|l|l}
\hline
${\rm
MODEL}$&${\sigma_8}$&${h}$&${\Omega_0}$&${\Omega_{\Lambda}}$&${\Gamma}$\\
\hline
${\rm SCDM}$ & ${0.67}$ & ${0.5}$ & ${1.0}$ & ${0.0}$ & ${0.5}$ \\
\hline
${\rm OCDM}$ & ${0.85}$ & ${0.7}$ & ${0.3}$ & ${0.0}$ & ${0.21}$ \\
\hline
${\rm \Lambda CDM}$ & ${0.91}$ & ${0.7}$ & ${0.3}$ & ${0.7}$ & ${0.21}$ \\ 
\hline
${\rm \tau CDM}$ & ${0.51}$ & ${0.5}$ & ${1.0}$ & ${0.0}$ & ${0.21}$ \\
\hline
\end{tabular}
\end{table}

\subsection{Matching the luminosity function of high-redshift galaxies}  

Steidel and collaborators (Steidel \& Hamilton 1992; Steidel, Pettini
\& Hamilton 1995; Steidel et al.~1996; Giavalisco, Steidel \&
Macchetto 1996) detected a population of actively star-forming
galaxies at redshift 2.5 to 3.5 with a space density of about $10^{-2}
h^{3} \Mpc^{-3}$ --- similar to that of present-day galaxies --- and
star formation rates of order $1-100\,M_{\odot}\,\yr^{-1}$. 
The optical observations carried out correspond to the rest-frame UV and
probe the star formation rate rather than the integrated stellar
light. Little is known at present about the masses of these objects
and the relation of their mass to the rate of detected star formation.
As Figure 1 demonstrates,  the  space density of these star-forming galaxies
corresponds to those of haloes with masses of $~10^{12.5}\,M_{\odot}$ 
and virial velocities of $~300\,\kms$ (see also Baugh et al.~1997).
Further evidence for masses of this order comes from the  
strong clustering of these galaxies (Steidel et al.~1997,   Bagla
1997, Jing \& Suto 1997, Frenk et al.~1997, Peacock 1997). 
A reasonable fit to the luminosity function of 
these objects can  be obtained  assuming a linear relation between
star formation rate and halo mass, i.e. a constant mass-to-light ratio. 
In Figure 2a we plot  such a  fit for the SCDM model. 
A weakly non-linear relation is also consistent with the
data and would indeed be required to match the  shallow slope of
the luminosity function at the high luminosity end reported recently by
Bershady et al. (1997). The fits  obtained for the other CDM variants
are of the same quality. 
Note, however, that the observed $H_{\beta}$ widths of these galaxies, 
$\sigma \sim 80\, \kms$ (Pettini et al.~1997), might be 
in conflict with the virial velocities of $300\, \kms$ quoted 
above.
\begin{figure*}
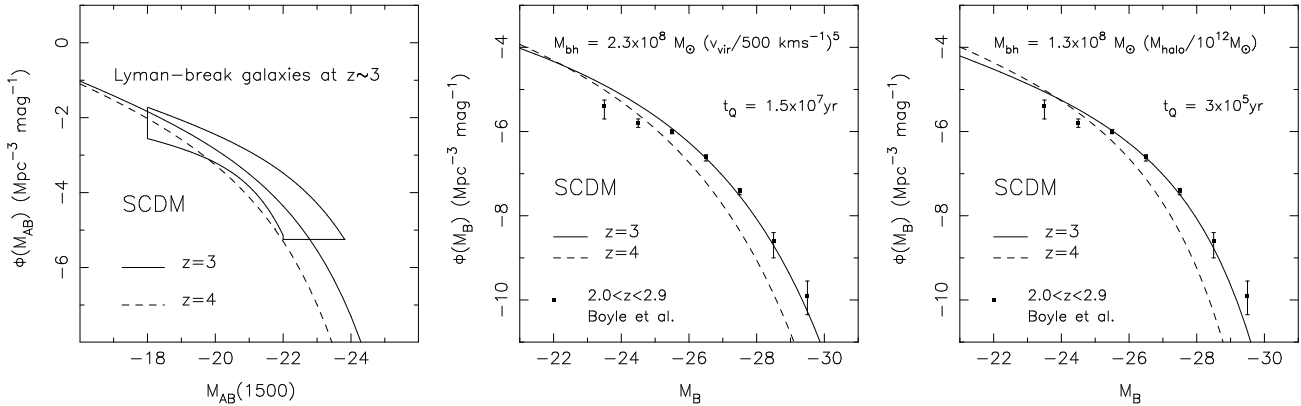

\centerline{
\hspace{0.cm}\psfig{file=haehnelt_fig2a.ps,width=5.5cm,angle=0.0}
\hspace{0.25cm}\psfig{file=haehnelt_fig2b.ps,width=5.5cm,angle=0.0}
\hspace{0.25cm}\psfig{file=haehnelt_fig2c.ps,width=5.5cm,angle=0.0}
}
\vspace{0.25cm}
\caption{{\it Left panel}: The luminosity function of star-forming galaxies 
at$z$=3. The magnitudes are AB magnitudes at $1500 \AA$.  
The model (indicated by the solid and dashed lines) assumes a linear 
relation between luminosity and halo mass.
{\it Middle panel}: The B-band QSO luminosity function at $z$=3
inferred from Boyle et al.~(1988).  
The positive term of the time derivative of the mass function in 
Figure 1 has been used to 
estimate the space density of newly-formed haloes.
QSOs were assumed to radiate with a light curve 
$L_{\rm B}(t) = f_{\rm B}\,f_{\rm Edd} \, L_{\rm Edd}\exp{(-t/t_{\rm
Q})}$, where $f_{\rm Edd}$=1 is the ratio of bolometric 
to Eddington luminosity and  $f_{B}$=0.1 is the fraction of the 
bolometric luminosity radiated in the B-band. The black hole mass is 
assumed to  be determined by the virial velocity of the DM halo, 
$M_{\rm bh}\propto v_{\rm halo}^{5}$ (model A) 
and the black hole mass indicated on the plot scales with 
$(f_{\rm Edd} f_{\rm B}/0.1)^{-1}$. {\it Right panel}: 
same as middle panel but for a shorter lifetime and a linear relation 
between black hole and halo mass,  $M_{\rm bh}\propto M_{\rm halo}$ 
(model B). \label{fig_2}}
\end{figure*}

\subsection{Matching the luminosity function of optically selected 
QSOs at high redshift}  

The space density of optically selected QSOs at $z$=3 with 
$M_{\rm B} <-23$  is smaller than that of the detected star-forming 
galaxies by a factor of a few hundred. It is therefore 
less obvious how to match the mass function 
of DM haloes with the B-band luminosity function of QSOs. 
As demonstrated 
by HR93 the well-synchronized evolution of optically selected 
QSOs can be linked  to the  hierarchical growth of DM haloes on a
similar timescale if the duration $t_{\rm Q}$ of the optically bright
phase is considerably shorter than the Hubble time. The comoving space
density of  DM haloes hosting QSOs then exceeds that of 
observed QSOs by a factor $1/t_{\rm Q}\, \Omega^{1/2}\,
H_{0}\,(1+z)^{5/2}$ where we have assumed that one  new generation of 
haloes forms per unit redshift. For  small $t_{\rm Q}$ this 
comes more and more in line
with the predicted space density of DM haloes and that of 
star-forming galaxies at high redshift. However, hardly anything 
is known  about the masses of the host objects of optically selected
QSOs and this still leaves considerable  freedom in the exact choice 
of $t_{\rm Q}$.  Following the approach of HR93 we 
estimate the formation rate of active black holes by 
taking  the positive term of the time derivative of the halo mass
function and a simple  parameterization for the black 
hole formation efficiency. 
It is further assumed that active black holes radiate  with a light 
curve of the form, $L_{\rm B}(t) = f_{\rm B}\,f_{\rm Edd} \, 
L_{\rm Edd}\exp{(-t/t_{\rm Q})}$, where
$f_{\rm Edd}$ is the ratio of bolometric to Eddington luminosity
and  $f_{B}$ is the fraction of the bolometric 
luminosity radiated in the B-band.

We  obtain reasonable fits for a wide range of lifetimes 
and for all the  CDM variants if we allow ourselves some freedom in 
the relation between halo mass and black hole mass.  
There are, however, systematic trends: with 
increasing lifetime the black hole mass has to
become a progressively more nonlinear function of the halo mass and
the black hole formation efficiency has to  decrease  in order to match
the luminosity function of QSOs. This is due to the fact that QSOs are
identified with rarer and more massive haloes with increasing
lifetime, and these fall on successively steeper portions of the halo mass
function.  In Figure 2b and 2c we plot two specific choices of parameters
which we denote as model A and B hereafter. For Fig. 2b a QSO lifetime
close to the Salpeter timescale $t_{\rm Salp}\,= 
\epsilon \sigma_{\rm T} c/4\pi G m_{\rm p} = \,4.5\,
\epsilon_{0.1}\, 10^{7}\yr$ and a scaling of black
hole mass with halo virial velocity as $M_{\rm bh}\,\propto\, v_{\rm
halo}^{5}\propto M_{\rm halo}^{5/3} (1+z)^{5/2}$ was assumed 
($\epsilon$ is the total efficiency for transforming accreted rest mass 
energy into radiation).  A physical motivation for this particular 
dependence is discussed in section 3. 
Fig. 2c shows the case of a linear relation
between the halo and black hole mass advocated by Haiman and Loeb
(1997,HL97) which requires a QSO lifetime of less than $10^{6}\yr $ 
-- much shorter than the Salpeter time  for usually assumed 
values of $\epsilon$. 
Figure 2 is for the SCDM model while Figure 3 is for the other CDM 
variants (model A only). In principle $t_{Q}$ could also depend 
on mass or other parameters and the degree of agreement which can
be obtained with our  simple assumptions seems gratifying.

\begin{figure*}
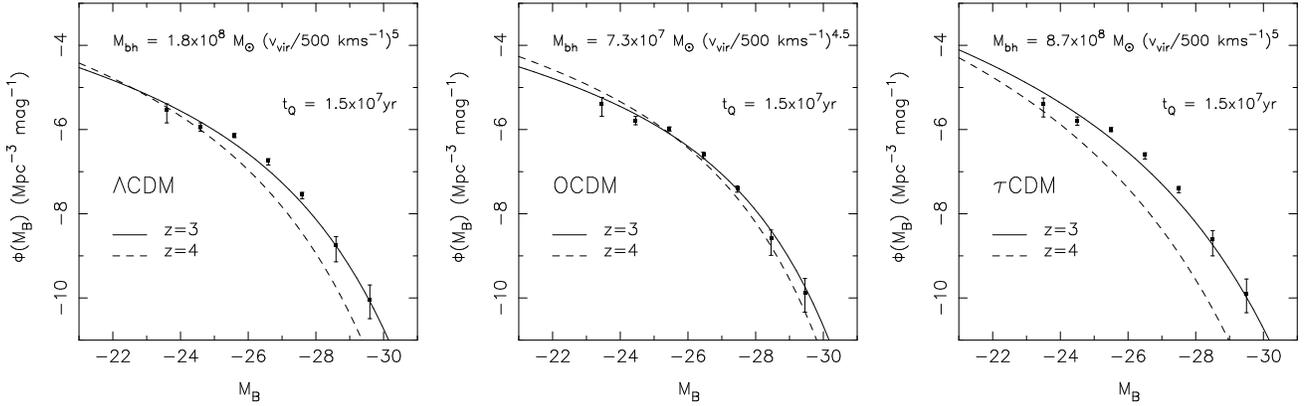

\centerline{
\hspace{0.cm}\psfig{file=haehnelt_fig3a.ps,width=5.5cm,angle=0.0}
\hspace{0.25cm}\psfig{file=haehnelt_fig3b.ps,width=5.5cm,angle=0.0}
\hspace{0.25cm}\psfig{file=haehnelt_fig3c.ps,width=5.5cm,angle=0.0}
}
\vspace{-0.0cm}
\caption{The B-band QSO luminosity function at $z\,=\,3$. 
Same as middle panel of Fig.2 but for three different 
CDM variants. \label{fig_2}}
\end{figure*}

\subsection{The demography of black holes in nearby galaxies}  

The last few years have seen tremendous progress  in establishing 
the existence of supermassive black holes in  our own 
Galaxy and nearby galaxies.   There are now a number 
of excellent cases (including that of
our own Galaxy)  where  observations strongly imply 
the presence of a relativistic potential well
(Watson \& Wallin 1994; Miyoshi et al.~1995; Genzel et al.~1997).
Furthermore, the number of mass estimates has increased to a level where it 
becomes possible to  make statistical arguments.  Magorrian et al. 
(1997; Mag97 henceforth) published a sample of about thirty  estimates for the
masses of the putative black holes in the bulges of nearby galaxies. 
Mag97 confirm previous claims of a strong
correlation between bulge and black hole mass (Kormendy \& Richstone
1995, but see Ford et al.~1997).  A linear relation of the form, 
$M_{\rm bh}= 0.006\,M_{\rm bulge}$, was obtained by Mag97 as a 
best fit. However, considering the large scatter  
a mildly non-linear relation  would probably also be consistent with 
the data. We would further like to note here that a linear relation 
between black hole to bulge mass does not necessarily imply a linear relation
between black hole and halo mass and as we will argue 
later a non-linear relation might be more plausible.
Fugukita, Hogan \& Peebles (1997) estimate the total mass density 
in stellar bulges  as 
$ 0.001h^{-1} \le \Omega_{\rm bulges} \le 0.003h^{-1}$ and together  with 
the above ratio of black hole to bulge mass  we get, 
\begin{eqnarray}
\lefteqn{\rho_{\rm bh} ({\rm nearby\ galaxies})=}\qquad\qquad\qquad&&
\nonumber\\
3.3h\times 10^{6} 
&\left (\frac{M_{\rm bh}/M_{\rm bulge}}{0.006}\right )\,
\left (\frac{\Omega_{\rm bulge}}{0.002h^{-1}}\right ) \,
M_{\odot}\,\Mpc^{-3}.&  
\end{eqnarray}
Considering  the complicated  selection biases of the Mag97 sample,
the  small sample size and possible systematic errors 
in the black hole mass estimates this number 
is still  rather uncertain. Van der Marel (1997) e.g.
emphasizes the sensitivity of black hole mass estimates to the
possible anisotropy of the stellar velocity distribution and argues 
that the Mag97 mass estimates  might be systematically too high.

Nevertheless, as pointed out by Phinney 
(1997; see also Faber et al.~1996)  $\rho_{\rm bh} ({\rm nearby\ galaxies})$ 
exceeds the mass density in black holes needed to explain the 
blue light of QSOs purely by accretion onto super-massive black holes,  
\begin{equation}
\rho_{\rm acc} ({\rm QSO}) = 1.4\times 10^{5}\,
\left (\frac{f_{\rm B}\,\epsilon}{0.01}\right )^{-1}\,
M_{\odot}\,\Mpc^{-3},
\end{equation}
by a factor of  about ten  unless  the value of $f_{\rm B}\, \epsilon$ 
is smaller than usually assumed (Soltan 1982, Chokshi \& Turner 1992). 
While  a few years ago it seemed difficult to discover  
the total mass in black holes necessary to explain the blue light 
emitted by QSOs at high redshift, black hole detections in nearby
galaxies  now suggest  that accretion onto supermassive black holes  
may actually be rather inefficient in producing blue light.

\section{The formation and accretion history of supermassive black holes}

\subsection{When and how did  supermassive  black holes gain most of
their mass?} 

There are three options to explain the apparently  large value 
of  $\rho_{\rm bh} ({\rm nearby\ galaxies})/\rho_{\rm acc} 
({\rm QSO})$: 
(i) $\rho_{\rm bh} ({\rm nearby\ galaxies})$ is strongly
overestimated, or 
(ii) $f_{B}\,\epsilon$ 
during the optically bright phase 
is smaller than previously assumed, or 
(iii) supermassive black holes do not gain most of their mass during 
the optically bright phase.  A plausible solution with   
$f_{B}\,\epsilon$ significantly smaller than 0.01  
is discussed in the next section. Here we will explore the 
last possibility somewhat further.  The typical mass of a black hole at
the end of the optically  bright phase of duration $t_{\rm Q}$ exceeds that 
accreted during this phase by a factor  
$M_{\rm bh}/M_{\rm acc} = f_{\rm edd}^{-1}\, t_{\rm Salp}/t_{\rm Q}$.
This factor should be larger than 1 and smaller than 
$\rho_{\rm bh} ({\rm nearby\ galaxies}) /\rho_{\rm acc} ({\rm QSO})$
and  therefore,
\begin{eqnarray}
1&\le &f_{\rm Edd}^{-1}\,\epsilon^{-1}\,\left (\frac{t_{\rm Q}}{4.5\times
10^{8}\yr}\right ) ^{-1} \nonumber\\
&\le  &{25h \left ( \frac{f_{B}\,\epsilon}{0.01}\right)\,\left ( \frac{\rho_{\rm bh} ({\rm nearby\ galaxies})}{3.3h\times 
10^{6}\, \Mpc^{-3}}\right )} .
\end{eqnarray}

The question when supermassive black holes
gained  most of their  mass is therefore closely related to  
$t_{\rm Q}$ and $f_{\rm Edd}$.  For bright  quasars, 
$f_{\rm Edd}$ must be  $\ga 0.1$; otherwise  excessively massive  
individual black holes would be required. 
Furthermore, $f_{\rm Edd}$ will always be smaller than unity
even if the ratio of the  accretion rate to that necessary to 
sustain the Eddington luminosity, $\dot m$, greatly 
exceeds unity. This is because a ``trapping
surface'' develops at a  radius proportional to $\dot m$, 
within which the radiation advects inwards rather than escapes.
In consequence, the emission efficiency declines inversely with 
$\dot m$ for $\dot m >1$ (Begelman 1978).

For $0.1 \le f_{\rm Edd} \le 1$  the possible range 
for $t_{\rm Q}$  is, 
\begin{eqnarray} 
\lefteqn{2h^{-1}\times 10^{6}\left  
(\frac{f_{\rm B}}{0.1} \right )^{-1} \, 
\left ( \frac{ \rho_{\rm bh} ({\rm nearby\ galaxies})}{3.3h\times
 10^{6} \, \Mpc^{-3}}\right )^{-1}\yr}&&\nonumber\\
&\qquad\qquad\le  t_{\rm Q}& \le 
4.5 \times 10^{8}\, \left (\frac{\epsilon}{0.1}\right)\, \yr.  
\end{eqnarray}  

If $t_{\rm Q}$ is very short (as in model B) 
and  $f_{B}\,\epsilon$ is not significantly smaller than 0.01
it seems  inevitable that
supermassive black holes have acquired most of their mass  
{\it before} the optically bright phase. We would like to point out 
here that a value of $\rho_{\rm bh}$ as large or larger than  we 
infer from Mag97  is actually needed for small  
$t_{\rm Q}$ ($\la 10^{6} \yr$).

For the remainder of this section we assume that $t_{\rm Q}$ is of order
the Salpeter time (as in model A).  The ratio of accreted mass 
to total mass at the end of the optically bright phase 
is then equal to $f_{\rm Edd}^{-1}$. 
If $f_{\rm Edd} \sim 1$  during the optically bright phase 
(and if  $f_{B}\,\epsilon$ is not significantly smaller than 0.01)
then the corresponding gain in mass by a 
factor $\rho_{\rm bh} ({\rm nearby\ galaxies})/\rho_{\rm acc} ({\rm QSO})$
indicated by Mag97 has to occur {\it after} the optically 
bright phase.  As the accretion should 
not be optically bright the most plausible options are 
advection dominated accretion flows 
(ADAFs) and dust-obscured accretion (Narayan \& Yi 1995, Fabian et
al.~1997 and earlier references cited therein). ADAFs require low 
accretion rate with $\dot m < m_{\rm crit}$ where  
$\dot m_{\rm crit} = 0.3\alpha_{\rm ADAF}^{2}$ and $\alpha$ is the 
Shakura-Sunyaev viscosity parameter.  
There  is therefore a maximum growth factor for the black hole 
mass density  due to ADAFs $\sim  3.0\alpha_{\rm ADAF}^{2} 
t_{\rm ADAF}/t_{\rm Salp}(\epsilon = 1)$  
and  \hfill \break
$\alpha_{\rm ADAF}\ga 0.3\,[\epsilon\, \rho_{\rm bh} 
({\rm nearby\ galaxies})/\rho_{\rm acc} ({\rm QSO})]^{0.5}$
would be required even if the accretion lasts all the way 
from $z$=3 to $z$=0. 

If, however, $f_{\rm Edd} \sim 0.1$  
(and $t_{\rm Q} \sim t_{\rm Salp}$)then the gain in mass by a
factor  $\rho_{\rm bh} ({\rm nearby\ galaxies})/\rho_{\rm acc} 
({\rm QSO})$ has to  occur  {\it before} the optically bright
phase as in the case of small $t_{\rm Q}$.  
For ADAFs this would require $\alpha_{\rm ADAF} \sim 1$ and is therefore
hardly plausible. In this case  dust-obscured accretion  would be the only
viable option.

\subsection{How and when did supermassive black holes first form?}  

A clue to this question might lie in the fact that no black holes 
smaller than $10^{6}\Msol$ have been detected and that there  might be a 
physically determined lower limit to the mass of a supermassive 
black hole. A natural lower bound would be  a few times $10^{5} \Msol$  
the upper limit for which a supermassive  star can 
ignite hydrogen before it undergoes the post-Newtonian instability
(see Bond, Carr \& Arnett 1984 and references therein).
To see this  imagine the following scenario. The gas in a newly  
collapsing halo (which has not yet acquired a central black hole) 
settles into a  self-gravitating disk.  
The rotational velocity of the self-gravitating 
disk  can  be related to the virial velocity of the halo 
$v_{\rm halo} = \sqrt{GM_{\rm halo}/r}$ as follows, 
\begin{equation} 
v_{\rm rot} =   j_{\rm d}^{-1} \,   
\left( \frac {\lambda}{0.05} \right )^{-1} \, 
\left (\frac {m_{\rm d}}{0.1} \right ) \,  v_{\rm halo} ,
\end{equation} 
where $\lambda=JE^{0.5}G^{-1}M^{-2.5}$ is the usual angular 
momentum parameter characterizing the halo, $m_{\rm d}$
the ratio of disc to  halo mass and $j_{\rm d}$ the ratio  of 
disc to  halo specific angular momentum 
(see also Mo, Mao \& White 1997).  For an isothermal DM halo 
$j_{\rm d}^{-1} \, \left (\frac {\lambda}{0.05} \right )^{-1} \, 
\left( \frac {m_{\rm d}}{0.1} \right )\ge 1$ is required for the disc 
to be self-gravitating (for flatter central profiles, self-gravitation
is even more likely). The self-gravitating disc  will be subject 
to violent gravitational instabilities (Shlosman, 
Begelman \& Frank 1990). For simplicity let us 
assume that the rotation velocity 
is constant with radius.  The rate of accretion due to gravitational 
instabilities can then be written as $\dot M_{\rm acc} 
= \beta v_{\rm rot}^{3}/G$ which is also constant  with radius 
and where we have defined $\beta$ as the  ratio of 
dynamical to accretion timescale. Rather small values of $\beta \sim
0.001$ are already sufficient for appreciable accretion rates
if the rotational velocity of the disk is as large, or larger than, 
the virial velocity of the DM halo. 
The first $10^6 \Msol$ can be transported to the centre 
on a time scale shorter than $10^{6} \yr$ (the lifetime of 
hydrogen burning star  radiating at the Eddington limit) 
if $v_{\rm rot}> 200 \,(\beta/0.001)^{-1/3} 
\, \kms $.  The supermassive star will then collapse due to the 
post-Newtonian instability before it  runs out of fuel.
The  accretion rate onto the newly-formed black hole 
will exceed its  Eddington accretion rate at first
by a factor $100\, (\beta/0.001)\, (v_{\rm rot}/200\kms)^{3}$. 
In this phase the  object will radiate at the Eddington limit and 
the luminosity increases linearly.

The subsequent evolution will depend on whether the emitted energy can 
back-react on the accretion flow prior to  fuel exhaustion. 
Such a back-reaction ought to   occur 
when the luminosity exceeds  the energy  deposition rate   
necessary to unbind the disc on a dynamical timescale
which scales as $v_{\rm rot}^{5} /G$. 
This will limit the black hole mass to, 
\begin{eqnarray}
M_{\rm bh} &\sim& 10^8\Msol\,  (f_{\rm kin}/0.001)^{-1}\,
 j_{\rm d}^{-5} \,   
\left( \frac {\lambda}{0.05} \right )^{-5} \, 
\left (\frac {m_{\rm d}}{0.1} \right )^{5} \, \nonumber\\ 
&&\qquad \qquad \times
\left (\frac {v_{\rm halo}}{400\, \kms}\right )^{5} \Msol,
\end{eqnarray}
where $f_{\rm kin}$ is the fraction of the accretion luminosity 
which is deposited as  kinetic energy  into the accretion flow
(cf. Silk \& Rees 1998). The accretion rate will then 
drop dramatically. The back-reaction timescale  will 
be related to the dynamical timescale of the outer parts 
of the disk and/or the core of the DM halo  
and should  set the  duration of the optically bright phase. 
It is interesting to note here that the accretion rate  will change  
from super-Eddington to  sub-Eddington without much gain in mass 
if the back-reaction timescale is  shorter than the Salpeter time. 
The overall emission efficiency is than 
determined by the value of $\dot m$ when the back-reaction sets in
and is reduced by a factor  $1/\dot m$ compared to accretion 
at below the  Eddington rate.  This results in a low overall
value of $f_{B}\,\epsilon$  which is -- as we  discussed  in the previous 
section -- one option to explain a large value of $\rho_{\rm bh} 
({\rm nearby\ galaxies})/\rho_{\rm acc}({\rm QSO})$. 
Subsequently the accretion rate will fall below 
the critical rate permitting  an ADAF, and the black hole 
will spend most of its time accreting with low efficiency.
As discussed above it will depend on the value of $\alpha_{\rm ADAF}$ 
whether black holes can gain most of their mass during this late 
phase of accretion. 

\begin{figure}
\vspace{-0.5cm}
\centerline{
\hspace{0.0cm}\psfig{file=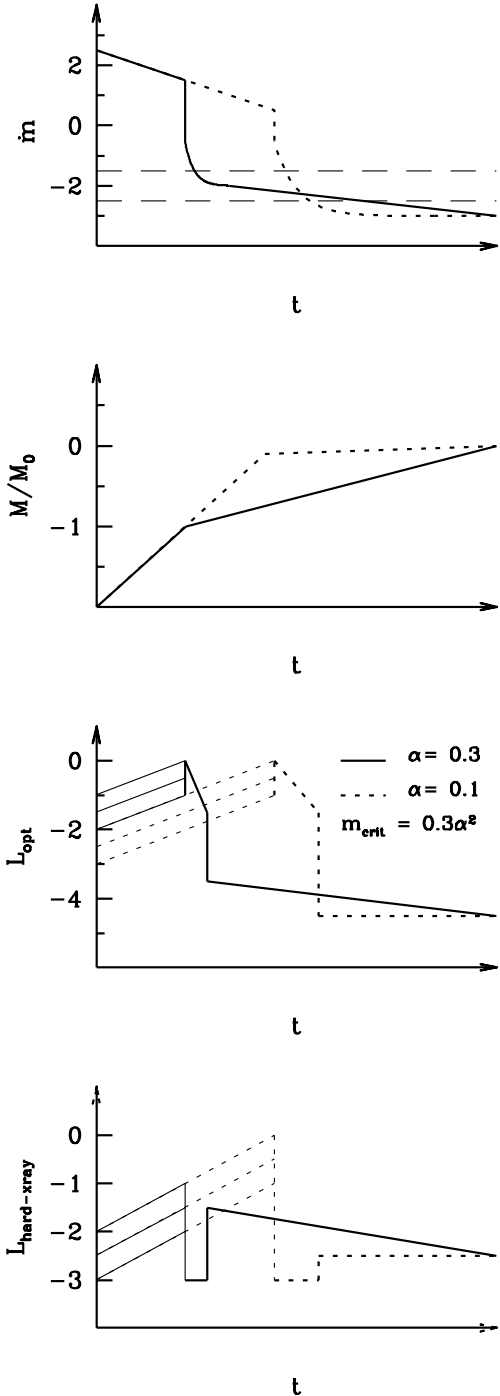,width=7.0cm,angle=0.0}
}
\vspace{-0.0cm}
\caption{Two accretion histories with low overall 
efficiency for producing blue light. 
The solid curves describe an accretion history 
where  most of the mass is accreted during a late 
and prolonged  ADAF phase while the 
dashed curve is a case where the black hole gains most of its 
mass during a short-lived early phase with $\dot m>1$. The panels 
show (from top to bottom) mass  accretion rate
in terms of the Eddington accretion rate the mass relative to the
final mass and the optical  and hard X-ray luminosity. 
The dashed lines in the top panel indicate  the critical 
accretion rate for $\alpha$=0.1 and $\alpha$=0.3. 
The spectral energy distribution  for accretion  with  
$\dot m>1$ is rather uncertain (as indicated by the three parallel 
lines  for $\dot m > 1$ in the two bottom panels) 
and should depend on the absorbing column and the dust content 
of the outer parts of the self-gravitating
disc and the host galaxy.\label{fig_3}}
\end{figure}

As is apparent from equation (6) the black hole mass is  very 
sensitive to the value of  $m_{\rm d}/j_{\rm d}\lambda$.
Numerical simulations have demonstrated that the distribution of $\lambda$  
depends very little on any halo properties (Barnes \&
Efstathiou 1987; Lemson \& Kauffmann 1997). Numerical 
simulations, however,  also show that angular momentum is 
very efficiently transferred  from the gas to the DM halo unless 
considerable energy and/or momentum input prevents the infalling gas 
from clumping (Navarro \& Steinmetz 1997). 
A moderate change in the ability of the gas to transfer angular 
momentum to the DM halo and in the amount of gas 
settling into a self-gravitating disc with redshift
would   result in a strong redshift dependence of the black 
hole formation efficiency. 
Such a strong redshift dependence was proposed by HR93 to explain 
the rapid decline of the typical QSO luminosity 
at $z<3$.  As discussed by Cavaliere \& Vittorini (1997) the decrease of 
the  ability of the gas to cool  once the hierarchical 
build-up of DM haloes has progressed from galaxy to 
group and cluster scales and the reduced  merger rates of galaxies 
compared to that of DM haloes contribute to this rapid decline.
The redshift 
evolution of the black hole formation 
efficiency and  the  expected  scatter in  $\lambda$ 
can  easily explain the  large scatter in the 
observed black hole to bulge mass relation. Furthermore, 
it seems likely that the bulge-to-disk ratio of a 
galaxy is also related to the value of $m_{\rm d}/j_{\rm d}\lambda$. 
This fits in well with the fact that the black hole mass 
seems to correlate  better with the bulge mass than the total 
mass of the galaxy and naturally explains the correlation 
of black hole formation efficiency with the Hubble type
of the galaxy. 
Obviously the evolution described in this section need not to occur 
strictly in that time sequence. In hierarchical cosmogonies, where DM 
haloes undergo continuous merging the process will
be re-started several times. At which step of the above scheme the 
evolution starts depends on whether the accretion rate 
in the core of
a newly-formed DM halo on a given level of the  hierarchy is 
super-Eddington, sub-Eddington or sub-critical with respect 
to the supermassive black hole already present.  
A question which  we have not addressed here but that certainly 
deserves  more attention is the merging history of supermassive 
black holes. Here we would only like to remark that 
in models with a strictly linear relation of black hole 
to halo mass constant with time (as in model B) 
the growth of black holes will actually be dominated by the  
merging of black holes, while a non-linear relation 
(as in model A) leaves room for a substantial gain in mass by 
accretion of gas at each step of the hierarchy.

\subsection{Possible accretion histories with low optical efficiency}

Figure 4 sketches two possible  accretion histories 
with a  low overall efficiency for producing 
blue light. The solid curves describe an accretion history 
where  most of the mass is accreted during the ADAF phase while the 
dashed curves are for an accretion history where the black hole gains
most of its mass during a short-lived early phase with 
with $\dot m>1$. The panels show (from top to bottom) mass  
accretion rate in terms of the Eddington accretion rate, the mass 
relative to the final mass and the optical  and hard X-ray luminosity. 
The accretion rate is constant at  the beginning with 
$\dot m >1$. The mass is therefore linearly rising and 
$\dot m$ decreases. The spectral energy distribution  for accretion  with  
$\dot m>1$ is rather uncertain (as indicated by the three parallel 
lines  for $\dot m > 1$ in the two bottom panels) 
and should depend on the absorbing column and the dust content 
of the outer parts of the self-gravitating disc and/or the host galaxy.
The sharp drop of $\dot m$ marks the onset of 
the back-reaction on the accretion flow and either 
the start or the peak of the optical bright phase 
(with a rather inefficient production of hard X-rays).
Once the accretion rate has fallen below the  critical rate 
for an ADAF (indicated by the dashed lines in the top panel) the
spectral energy distribution will change to one peaked in the 
hard X-ray waveband.

\section[]{Observational implications}

\subsection{Masses, luminosities and clustering properties of 
QSO hosts at redshift three}

\begin{figure*}
\centerline{
\hspace{-1.0cm}\psfig{file=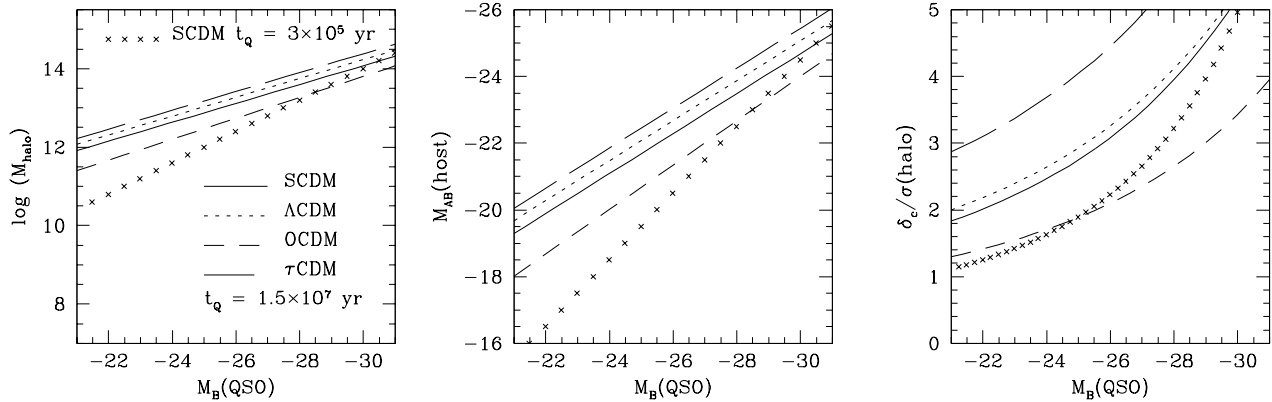,width=20.0cm,angle=0.0}
}
\vspace{-13.0cm}
\caption{Inferred properties of QSO hosts at $z\,=\,3$.
The curves are for model A ($M_{\rm bh} \propto v_{\rm halo}^{5}$)  
while the crosses are for model B ($M_{\rm bh} \propto M_{\rm halo}$,
SCDM only).{\it Left panel}: halo mass vs observed B magnitude of the QSO.
{\it Middle panel}: Host luminosity (AB magnitude at $1500 \AA$
against  observed B magnitude of the QSO.
{\it Right panel:} Peak significance. 
\label{fig_5}}
\end{figure*}

In Figure 5 we plot  some properties of the host objects predicted by our
fiducial models for the different CDM variants.   As expected
in model A,  where $t_{\rm Q}$ is longer , the halo masses are 
considerably larger than in model B especially at low luminosities. 
This is also reflected in the expected luminosity of the host galaxy. 
While in model A QSO
hosts ($M_{B}\,<\,-23$) would sample only the bright end of the
luminosity function of star-forming galaxies ($M_{AB}\,<\,-21$), in
model B the faintest QSOs would reside in objects well beyond the
current spectroscopic limit at these redshifts (Fig. 3b).  
Large $t_{\rm Q}$ seem therefore to stand a better chance to be 
consistent with the  large host galaxy luminosities of bright 
QSOs reported  by Aretzaga, Terlevich \& Boyle (1997). Figure 3c 
illustrates the ``rareness'' of host haloes/galaxies in terms of the
peak height in a Gaussian distribution. In model A QSO hosts
correspond to significantly rarer peaks.  The clustering strength 
of haloes  depends mainly on the rareness of the peaks
(Kaiser 1984, Mo\& White 1996, Bagla 1997) and  the two  models should  
therefore differ strongly in the predicted  clustering length of QSOs.
With the upcoming larger surveys such as the 2dF and the SDSS it 
should be possible  strongly to  constrain the 
clustering length,  and thus also  $t_{\rm Q}$.
The  strong clustering    recently confirmed by  
La Franca \& Cristiani (1997) seems  already to dis-favour  
small  $t_{\rm Q}$.

\subsection{Faint X-ray sources and the hard X-ray background}

As pointed out by many authors, the X-ray emission of optically 
selected QSOs  is too soft to explain the hard X-ray background.
Di Matteo \& Fabian (1996) and Yi \& Boughn (1997) argued  that  
the emission from  ADAFs has a spectral shape similar to  the hard 
X-ray background. Fabian et al.~(1997)  suggested 
that this might  also be true for 
dust-obscured accretion. It is therefore tempting  to link the 
rather large value of   $\rho_{\rm bh} ({\rm nearby\ galaxies})
/\rho_{\rm acc} ({\rm QSO})$ inferred from Mag97
to the origin of the  hard X-ray 
background and the recently  detected large space 
density of faint X-ray sources  (Almaini et al.~1996;
Hasinger et al.~1997, Schmidt et. al 1997,  McHardy et al.~1997, 
Hasinger 1998). The presence of extremely low-level 
optical AGN activity in a large fraction of galaxies 
reported by Ho, Fillipenko \& Sargent (1997)  would also fit 
in nicely with  such a picture. The present-day black hole 
mass density is sufficient
to produce the hard X-ray background if 
$f_{\rm hard X-ray}\, \epsilon \sim 0.002\,(\rho_{\rm acc}
({\rm hard\ X-ray})/10^{6} \Msol \Mpc^{-3})$. The efficiency of ADAFs is 
$\epsilon_{\rm ADAF} =0.1\, (\alpha/0.3)^{2}\, \dot m/\dot m_{\rm crit}$  
and decreases rapidly for small $\alpha$ and small 
$\dot m/\dot m_{\rm crit}$.  If the hard-Xray background was produced 
by  ADAFs onto supermassive black holes in ordinary galaxies
this requires a  value of $\rho_{\rm bh} ({\rm nearby\ galaxies})$ 
as high as we infer from Mag97, a large value of $\alpha$ 
and a value of $\dot m$ below but still close to  $\dot m_{\rm crit}$
lasting for a Hubble time for  the majority of all supermassive 
black holes. 

One should note here that at present it is not easy to 
discriminate observationally between  dust-obscured accretion and an 
ADAF as little work has been done on  the detailed spectral 
shape of an ADAF embedded  in an AGN environment. Furthermore, at 
faint flux levels and high redshifts a possible star-burst
contribution to the  total spectral energy distribution will 
become more and more important in the optical and probably also 
the soft X-ray.

\begin{figure*}
\centerline{
\hspace{0.0cm}\psfig{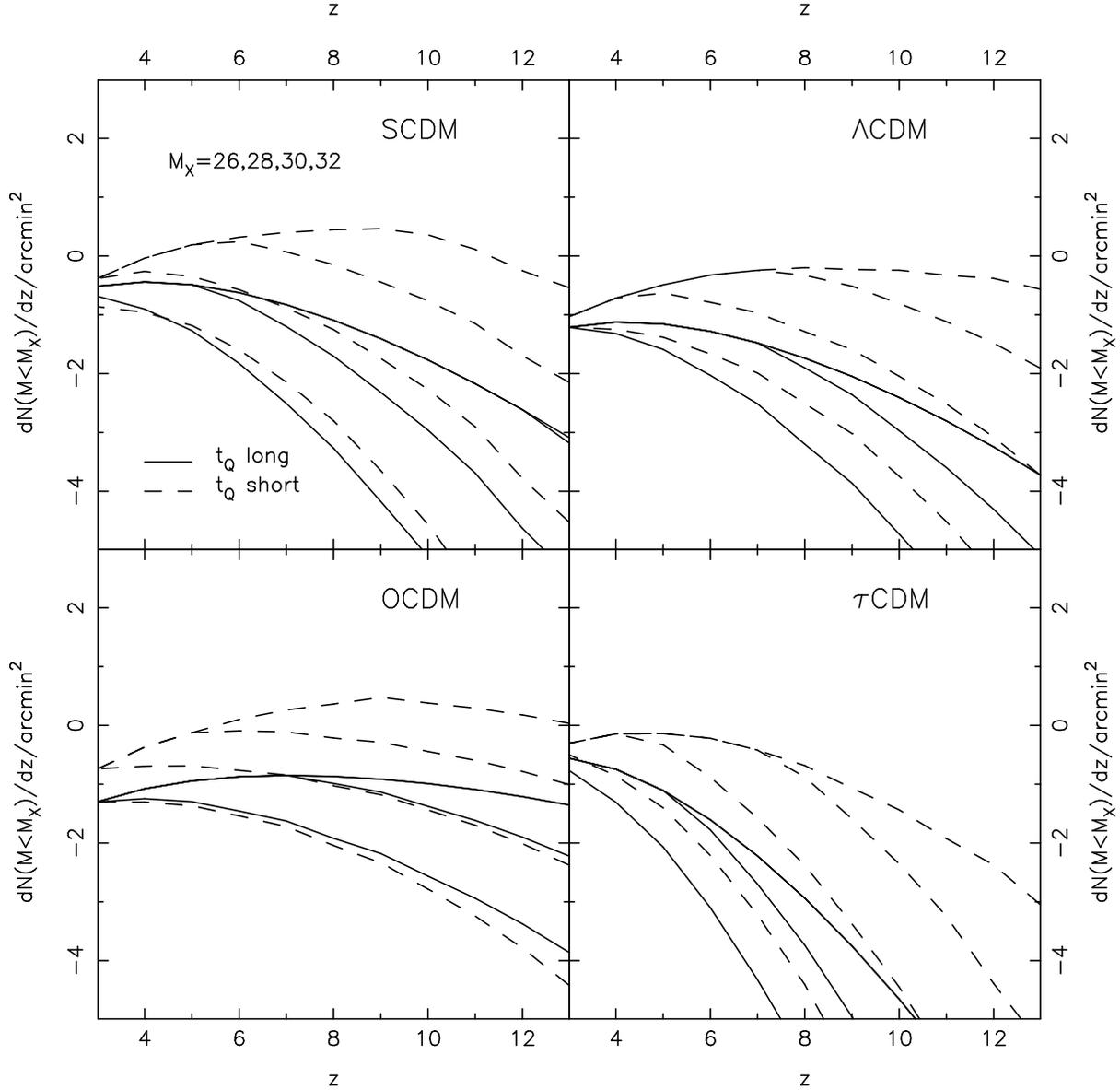}
}
\caption{Model predictions for the  number of faint point
sources per unit redshift  per arcmin$^{2}$ for the four 
different flux limits as indicated on the plot. 
The black holes  are assumed to emit a fraction $f_{\rm X}$=0.3 of 
the Eddington luminosity in the corresponding band (no extinction).
Model A ($t_{\rm Q}$ long) assumes a minimum black hole mass 
of $10^{6} \Msol$ and for both models  a minimum halo virial 
velocity $v_{\rm halo} \ge 30 \kms$  was assumed.
Note that the solid curves for $M_{\rm X}$=30 and $M_{\rm X}$=32 
coincide.}
\label{fig_6}
\end{figure*}

\subsection{Faint point sources in the optical and infrared at 
very high redshift}

In section 3.2 we  speculated that there might exist a minimum black
hole mass. Such a minimum mass will have a strong bearing on the 
number of point sources at faint magnitudes. 
Fig. 6 shows the predicted number of sources  
per unit redshift per arcmin$^{2}$
above a certain flux limit. We have assumed  that the sources 
emit a fraction $f_{\rm X}$=0.3 of the Eddington luminosity 
in the corresponding band.  No correction for dust absorption
or intervening \op absorption has been made.  Since the universe 
becomes optically thick to  \op absorption at $z\ga4$
these faint sources could only be detected longward 
of the redshifted \op wavelength (i.e K band or redder).
For model A a minimum black hole mass of $10^{6} \Msol$
and for both models  a minimum halo virial velocity 
$v_{\rm halo} \ge 30 \kms$  was assumed. At magnitudes 
$M_{\rm X} >28$ model A  predicts a  progressively  smaller number 
of faint sources than model B mainly due to the assumed  minimum 
black hole mass. As pointed out by HL97 the Next Generation Space 
Telescope (NGST) should be an excellent tool to probe the black hole
mass spectrum at high redshifts  and small masses. The rather 
small number of very faint point sources in the HDF might already be in 
conflict  with  a  linear relation of black hole  to bulge mass 
which extends below $10^{6} \Msol$ and to high redshift
(Madau, private communication; see also Almaini \& Fabian 1997).

\begin{figure*}
\centerline{
\hspace{0.0cm}\psfig{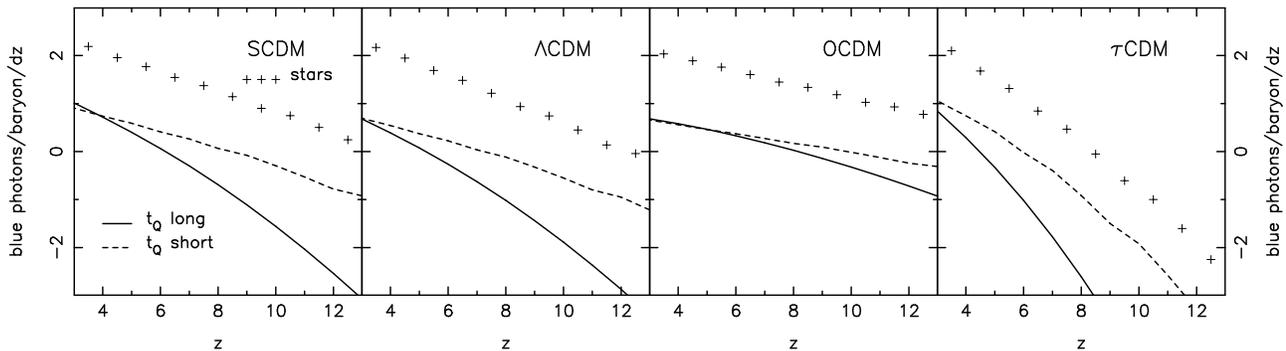}
}
\caption{Model predictions for the emission of blue light at $1500\AA$
due to quasars and stars. The number of photons emitted per baryon and
unit redshift are plotted (no extinction). The crosses show the
emission from star-forming galaxies. The solid line is the model with  
$M_{\rm bh} \propto v_{\rm halo} ^5$, and a minimum black hole mass of
$10^6\,M_{\odot}$ (model A). The dashed line is the model with 
$M_{\rm bh} \propto M_{\rm halo}$ (model B). A minimum halo
virial velocity of 30 $\kms$ was assumed. Star formation
rates are assumed to scale with $(1+z)^{1.5}$.}
\label{fig_7}
\end{figure*}

\section{The role of stars and QSOs for the re-ionization history of
the universe}

The contribution from newly-formed stars  relative  to that 
from  super-massive black holes to the re-ionization history of 
the universe is still
rather uncertain. At $z$=3 the flux of ionizing photons attributed
to observed QSOs is probably just sufficient to explain the observed
UV background at the hydrogen Lyman limit if the ionizing flux is as
low as indicated by recent studies of the flux decrement distribution
in QSO spectra which make use of hydrodynamical simulations (Rauch et
al.~1997; Bi \& Davidsen 1997). Rauch et al., however, find  that
the ionizing background necessary to fit the mean flux decrement
decreases more slowly towards higher redshift than the expected
contribution of observed QSOs. This indicates that at $z\,>\,3$ the
ionizing background might be dominated by stars. Direct estimates of the
stellar contribution to the ionizing background are still difficult as
we do not know how much ionizing flux is actually escaping
star-forming galaxies at high redshift. Just longward of the hydrogen
Lyman limit the photon flux due to star-forming galaxies exceeds that
due to quasars by a factor of about thirty.  The spectra of
star-forming galaxies certainly have a  strong intrinsic 
Lyman break but it is still unclear whether the ionizing flux 
from galaxies exceeds that due to QSOs even at $z\,=\,3$.  
A more precise assessment
of this question will has to await detailed spectra of star-forming
galaxies which extend beyond the hydrogen Lyman limit.  Theoretical
predictions for the relative contribution of stars and QSOs to the UV
background are also hampered by our lack of knowledge of the photon
escape fraction in both star-forming galaxies and QSOs, its possible
dependence on halo/galaxy properties and its evolution with
redshift. In Figure 7, we have plotted predictions for the photon
generation rate just longward of the hydrogen Lyman limit evolving our
two fiducial models backwards in time and assuming a minimum 
black hole mass of $10^{6} \Msol$ (units are photons per baryon
per unit redshift). In model A the black hole formation efficiency
strongly decreases with halo virial velocity. In such a model the
ionizing background should eventually be dominated by stars at high
redshift.  In model B, the fraction of ``blue'' photons coming from by
QSOs is only weakly redshift dependent. This is primarily due to the
linear scaling of the black hole and halo mass. As argued by
HL97, in such a model the ionizing background could be
dominated by QSOs at all redshifts.

Model predictions for the re-ionization redshift are even more
uncertain and as discussed by many authors they differ strongly
between different CDM variants and depend strongly on the assumed
escape fraction.  While mixed-dark matter models are just capable of
re-ionizing hydrogen at or before redshift five, in open models
hydrogen re-ionization by stellar radiation might occur well before
redshift ten.  Further clues to the question of how and when the
universe was re-ionized might be obtained from the thermal history of
the inter-galactic medium and the closely related issue of the
re-ionization history of helium. The spectrum of 
radiation emitted from star forming regions is generally believed 
to be too soft to fully ionize helium. If hydrogen were re-ionized by stars, 
then helium  would have to be re-ionized by the harder output from QSOs
which might have occurred considerably later. Recently, there has been
some  evidence for a change in the spectral shape of the
ionizing background at redshift three and only partial re-ionization
of helium at the same redshift has been reported (Songaila \& Cowie
1996, Reimers et al.~1997, but see Miralda-Escud\'e 1997 for a discussion
and also Boksenberg 1997). However, Haehnelt \& 
Steinmetz (1997) use an investigation of the observed Doppler
parameter distribution of QSO absorption lines to argue that helium
was probably fully re-ionized by UV radiation with a QSO-like spectrum
before $z$=4.

\section{Discussion and Conclusions}

The optical QSO luminosity function  at $z\sim 3$  can 
be plausibly matched with  the luminosity function  
of star forming galaxies at the same redshift  and the mass function 
of DM haloes predicted by a range of variants of CDM cosmogonies 
believed to comply with observational constraints in the 
low-redshift universe.  This is possible for 
lifetimes of optically bright QSOs  anywhere in the range 
$10^{6}$ to $10^{8}\yr$.  There is a correlation between
the lifetime and the required degree of non-linearity 
in the relation between black hole and halo mass.  
The non-linearity has to increase for 
increasing lifetime. Predicted  host halo masses, host galaxy 
luminosities,  and the clustering strength all increase with 
increasing lifetime and further observations of these offer 
our best hope of constraining  the duration of the optically bright 
phase of QSOs.  

The present-day  black hole mass  density  implied 
by the integrated luminosities of optically 
bright QSO may be significantly smaller than that inferred 
from recent black hole estimates in nearby galaxies
for generally assumed  efficiencies for producing blue light.
We have discussed  three possibilities for how and 
when this mass could be accreted in an optically inconspicuous 
way:
(i) in the early stages of accretion at   rates  
far above  the Eddington rate, 
(ii) by accretion where optical emission is obscured by dust, or 
(iii) in the late stages  of accretion at a  
rate  below the critical rate for an advection dominated 
accretion flow  with  an Shakura-Sunyaev parameter of 
$\alpha_{\rm ADAF}\ga 0.3$.
For the latter two  possibilities  the space density of faint X-ray 
sources should  exceed that  of optically selected QSOs 
considerably and  might reach that of present-day $L_*$ galaxies at 
the faintest flux levels. The expected shape of the luminosity
function of faint X-ray sources and its evolution with redshift
should, however, differ and improved 
determinations  of this luminosity function together with broader 
spectral coverage of the sources should soon determine to what extend 
one or more of these possibilities are realized.   

We have speculated that the formation and accretion history of supermassive 
black holes is determined by accretion at the centre of a gravitationally
unstable self-gravitating  disc in the core of a newly-formed 
dark matter halo and could proceed in the following steps:
\begin{itemize} 
\item[(i)] accumulation of enough material to form a supermassive star,  
\item[(ii)] growth to the limit for post-Newtonian  instability and collapse  
     to a supermassive black hole of $10^{6} \Msol$   
     on a   timescale less than  $10^{6} \yr$,
\item[(iii)] steady accretion  at well above 
      the Eddington rate  for about $10^{7} \yr$ with linearly 
      increasing mass,
\item[(iv)] reduction of the accretion rate due to back reaction on 
the accretion flow  once the luminosity 
exceeds  $v_{\rm rot}^{5}/G$ by a sufficient factor,  and 
\item[(v)] late accretion for a Hubble time at a (slowly decreasing) 
reduced level.  
\end{itemize}
This accretion history can explain the rather short optically bright 
phase, can account for a low overall efficiency for producing 
blue light and leaves room for the production of the hard X-ray background. 
It gives a physical explanation for the mass of the black hole formed 
and the predicted  scaling with virial velocity/mass  
of the halo that is  required  to match the optical QSO 
luminosity function  with the mass function of DM haloes 
for a duration of the optical bright phase of $10^{7} \yr$. 
The rapid redshift evolution of the black hole 
formation efficiency can be attributed to a modest evolution in the 
fraction of the gas settling into a self-gravitating  configuration 
at the centre of a DM halo and the ability of the gas to transfer 
angular momentum to the halo. 
It  might further have the interesting implication of a minimum
black hole mass of about $10^{6} \Msol$. The predicted  decreasing 
black hole formation efficiency  towards smaller halo masses and 
the presence of a minimum black hole mass may suggest that the ionizing 
background is  progressively dominated by stars at high redshift.

\section*{Acknowledgments}
We would like to thank George Efstathiou, Andy Fabian, 
Jordi Miralda-Escud\'e and Max Pettini for helpful 
discussions and comments on the manuscript.

\label{lastpage}

\end{document}